# GRAPHS : ASSOCIATED MARKOV CHAINS

Garimella Rama Murthy,
Associate Professor,
International Institute of Information Technology,
Gachibowli, HYDERABAD-500032

## ABSTRACT

In this research paper, weighted / unweighted, directed / undirected graphs are associated with interesting Discrete Time Markov Chains (DTMCs) as well as Continuous Time Markov Chains ( CTMCs ). The equilibrium / transient behavior of such Markov chains is studied. Also entropy dynamics ( Shannon entropy ) of certain structured Markov chains is investigated. Finally certain structured graphs and the associated Markov chains are studied.

## 1. Introduction:

Most natural / artificial dynamic phenomena are endowed with a non-deterministic description ( for their evolution in time ). Stochastic processes provide appropriate mathematical models of such phenomena. Among the class of stochastic processes, Markov chains are highly utilized for many dynamic phenomena because of demonstration of equilibrium behavior. Efficient computation of equilibrium / transient probability distribution of a Discrete time Markov Chain ( DTMC ) / Continuous Time Markov Chain ( CTMC ) is an ever-green research problem.

Given a homogeneous DTMC, there is a state transition matrix which is naturally associated with its equilibrium / transient behavior. Conventionally, the information provided by the state transition matrix is captured using the state transition diagram, which is a weighted directed graph. The weights are effectively probabilities in such a directed graph. In a similar spirit, a homogeneous CTMC is represented using a state transition rate diagram which is a weighted, directed graph. The weights are "state transition rates".

The author asked the following converse question.

Q: Can one or more DTMCs / CTMCs be naturally associated with a directed / undirected; weighted / unweighted graph ?

The above question is answered in the affirmative. Also some properties of such Markov chains are studied. This research paper is a culmination of such research efforts.

This research paper is organized as follows. In Section 2, unweighted graphs are considered and are associated with a homogeneous DTMC as well as homogeneous CTMC. Such Markov chains are studied for associated properties, particularly the equilibrium probability distribution. In Section 3, weighted graphs are considered and various DTMCs / CTMCs are associated with such graphs. In Section 4, dynamics of Shannon entropy associated with structured Markov chains is studied. In Section 5, certain interesting structured graphs and associated Markov chains are studied.

## 2. Unweighted Graphs : Associated Markov Chains:

- **Unweighted Graph: Discrete Time Markov Chain:**

Consider an unweighted, undirected graph with the symmetric adjacency matrix **A.** Let the associated diagonal matrix of degree values of each of the nodes be denoted by **D**. It is clear that

$$P = D^{-1} A$$

is clearly a stochastic symmetric matrix. Thus, we can naturally associate a homogeneous Discrete Time Markov Chain ( DTMC ) based on such a matrix. The state transition diagram of such a DTMC is obtained from the original graph by replacing each edge with two directed edges in opposite directions ( each edge corresponds to two directed edges ).

**Claim 1:** Since **A** is symmetric, it is clear that **P** is a doubly stochastic matrix.

**Claim 2**: It is well known that ( in the theory of DTMCs based on doubly stochastic state transition matrix ) the equilibrium distribution of such a DTMC is given by

$$\left[ \frac{1}{M}, \quad \frac{1}{M}, \ldots, \frac{1}{M} \right]$$

where the matrix **P** is of dimension M x M i.e. the equilibrium distribution achieves maximum entropy possible ( uniform distribution ). Also the transient probability distribution

$$\bar{\pi}(n+1) = \bar{\pi}(0) P^n \quad \text{for } n \geq 0.$$

is computed using the above equation.

**Unweighted Graph: Continuous Time Markov Chain (CTMC):**

Given a simple graph G with M vertices, its Laplacian matrix **L** is defined as

$$\mathbf{L = D - A},$$

where D is the diagonal degree matrix and A is the symmetric adjacency matrix. Now consider the following matrix

$$\mathbf{Q = - L = A - D.}$$

It is easy to see that **Q** satisfies all the properties required by a generator matrix. Thus **Q** is naturally associated with a homogeneous Continuous Time Markov Chain (CTMC). Furthermore let $\bar{e}$ be an all-ones column vector i.e.

$$\bar{e} = [\mathbf{1} \;\; \mathbf{1} \; \ldots . \mathbf{1}].$$

Then, it is easy to see that

$$\bar{e}^T Q = 0.$$

Hence the equilibrium probability distribution of such a homogeneous CTMC is given by

$$\left[\frac{1}{M}, \frac{1}{M}, \ldots, \frac{1}{M}\right] \quad \text{( Uniform Probability Mass Function ).}$$

It is well known from Information Theory that the above constitutes a maximum entropy probability mass function.

Naturally, we are interested in the transient probability distribution of such a homogeneous CTMC ( described by the generator matrix Q ). Such a Q is naturally associated with a state transition rate diagram ( a directed graph ). It is well known that the transient probability distribution is computed as :

$\bar{\pi}(t) = \bar{\pi}(0)\, e^{Q\,t} = \bar{\pi}(0)\, e^{(A-D)\,t}$.

In the above expression, matrix exponential $e^{Q\,t}$ can be efficiently computed since Q = A – D is a symmetric generator matrix.

- **Directed, Unweighted Graphs:**

  We now consider directed, unweighted graphs. Let $\boldsymbol{D_{in}}$ be the degree matrix associated with edges incident at various nodes and let $\boldsymbol{A_{in}}$ be the adjacency matrix associated with the edges incident at various nodes / vertices. As discussed above

$$Q_{in} = A_{in} - D_{in}$$

has the properties of a generator matrix**.** Thus, a homogeneous Continuous Time Markov Chain (CTMC) can be naturally associated with such a directed, unweighted graph. Since, the matrices $A_{in}$ , $D_{in}$ are symmetric, the matrix $Q_{in}$ is also symmetric.

- Hence the equilibrium probability distribution of such a CTMC is the maximum entropy probability distribution ( uniform distribution on M values ).

- Further the transient probability distribution can be computed as discussed previously.

Now let us consider arriving at the DTMC associated with directed, unweighted graph. It can be easily seen that

$$P = D_{in}^{-1} A_{in}$$

need not be defined as $D_{in}^{-1}$ may not exist for certain directed, unweighted graphs. We can overcome this problem in the following maner. If 'j'th diagonal element of the $D_{in}$ matrix is zero ( i.e. none of the directed edges is incident at the 'j'th vertex ), then by setting its inverse as equal to ONE, we can easily ensure that

$$P = D_{in}^{-1} A_{in}$$

is a stochastic matrix and thus a homogeneous DTMC can be naturally associated with it. In this ca se, it is easy to see that the adjacency matrix **A** neednot be a symmetric matrix. Hence P even though stochastic, neednot be doubly stochastic.

- Thus for the CTMC / DTMC associated with an unweighted, directed graph, the equilibrium / transient probability distribution can be computed using the standard procedure.

## 3. Weighted Graphs : Associated Markov Chains:

- Using the procedure discussed in Section 2, CTMC / DTMC can be naturally associated with such graphs ignoring the weights on the edges ( i.e. setting all the edge weights to one ). Such an approach can be utilized with weighted graphs that are directed or undirected. More interestingly, a DTMC can be associated with a weighted, undirected graph in the following manner:

- Consider any vertex, say $v_i$ . Let $v_i$ be connected to vertex $v_j$ . Normalize the weight of such edge by the sum of weights of all edges incident at the vertex $v_i$. $i.e.$

$$\overline{\boldsymbol{w_{ij}}} = \frac{w_{ij}}{\sum_k w_{ik}}$$ **i.e.** normalized weight from node $v_i$ to $v_j$.

It is thus clear that

$$\sum_j \overline{w_{ij}} = 1.$$

Thus, it is clear that the matrix of normalized weights is a stochastic matrix. Hence a Discrete Time Markov Chain (DTMC) is naturally associated with a weighted, undirected Graph. As discussed in the previous section, the equilibrium and transient probability distribution of such a DTMC can be determined. Details are avoided for brevity.

Now consider the following two matrices.

$$\overline{\boldsymbol{A_{ij}}} = \boldsymbol{W_{ij}}$$

i.e. A is the symmetric weight matrix. Also, let **'D'** be the diagonal matrix with the diagonal entries being the sum of weights of all edges incident at a vertex i.e.

$$\overline{D_{u}} = \sum_{j} w_{ij}.$$

Now define the following matrix

$$\bar{Q} = \bar{A} - \bar{D}.$$

It is easy to verify that the matrix $\bar{Q}$ has all the properties / structure of a generator matrix. Thus, a homogeneous CTMC is naturally associated with such a graph. As discussed previously, equilibrium / Transient probability distribution of such a CTMC can easily be determined.

Based on the approach discussed in Section 2, two pairs of Markov Chains ( one pair based on the edges coming into a node and the other pair based on the edges leaving a node ) can be naturally associated with a weighted directed graph. Details are avoided for brevity.

**Remark:**

In the research area of "symbolic dynamics", graphs ( directed / undirected ) are naturally associated with dynamical systems. It is expected that the results in this research paper can be combined with those from symbolic dynamics leading to new avenues.

- **Initial Probability Distribution ( Probability Mass Function ):**
  In the above discussion, various DTMCs/CTMCs are naturally associated with different types of graphs. The choice of initial probability distribution determines the evolution of transient probability distribution. In the following, we propose a specific initial probability mass function that could have important consequences. This initial probability distribution is based on the vertex degree distribution. Let $p_i$ be the probability that the Markov Chain starts in the 'i'th state ( $1 \leq i \leq M$ ). It is defined as

$$p_i = \frac{Degree\ of\ ith\ vertex}{Sum\ of\ Degrees\ of\ all\ vetices} \quad \text{for} \quad 1 \leq i \leq M.$$

- It should be noted that in the case of directed graphs, 'in-degree/out-degree' are naturally utilized to arrive at an initial probability distribution.

## 4. Graphs : Shannon Entropy Dynamics of Associated Markov Chains:

Let us consider an undirected, unweighted graph. We are interested in the dynamics of Shannon entropy associated with the transient probability mass function ( probability distribution ) of the associated Discrete Time Markov Chain ( considered in Section 2 ). The following Lemma is an interesting contribution in this direction.

**Lemma 1:** Consider an undirected, unweighted graph and the associated DTMC ( as discussed in Section 2 ). Let the DTMC be initialized with an arbitrary probability mass function ( distribution ). The Shannon entropy associated with the transient probability mass function ( time varying probability distribution ) is non-decreasing and reaches the maximum value ( corresponding to maximum entropy probability mass function on M values ).

**Proof:** The proof is based on the following result first derived by Feinstein in Information Theory [Ash, CoT].

- Let $B = [b_{ij}]$ be a doubly stochastic matrix i.e. $b_{ij} \geq 0$ $for\ all\ i,j$ and

$$\sum_{j=1}^{M} b_{i,j} = 1 \ for \ i = 1, \dots, M \ \ and \ \sum_{i=1}^{M} b_{i,j} = 1 \ for \ j = 1, \dots, M.$$

Given a set of probabilities $p_1, p_2, \dots, p_M$ , define a new set of probabilities $\overline{p_1}, \overline{p_2}, \dots, \overline{p_M}$ in the following manner:

$$\overline{p_i} = \sum_{j=1}^{M} b_{i,j} \ p_j \ \ for \ i = 1,2, \dots . M$$

Then, the Shannon entropy associated with the two probability mass functions satisfies the following inequality

$$H(\overline{p_1}, \overline{p_2}, \dots, \overline{p_M}) \geq H(p_1, p_2, \dots, p_M)$$

with equality if and only if probabilities ( $\overline{p_1}, \overline{p_2}, \dots, \overline{p_M}$ ) are a rearrangement of the probabilities $(p_1, p_2, \dots, p_M)$.

From Section 2, we have that the state transition matrix of associated DTMC is given by

$$P = D^{-1} A,$$

where A is the symmetric adjacency matrix. Hence P is symmetric and stochastic. Thus, P is necessarily a doubly stochastic matrix. From the theory of DTMCs, we realize that the transient probability distribution satisfies the following equation:

$$\bar{\pi}(n+1) = \bar{\pi}(n) P \ \ for \ n \geq 0,$$

where $\bar{\pi}(n)$ is the row vector of transient probabilities at time 'n'. Since P is doubly stochastic, using the Feinstein's result, we have that the Shannon entropy associated with the transient probability distribution is non-decreasing. Also, it is clear that the transient probability distribution converges to the equilibrium probability distribution $\left[ \frac{1}{M}, \frac{1}{M}, \dots\dots, \frac{1}{M} \right]$ ( i.e. the maximum entropy PMF ). Thus, the Shannon entropy is non-decreasing and reaches the maximum possible value in equilibrium.

Q.E.D.

- Similar results are being derived for DTMCs / CTMCs discussed in Section 2,3.

- **Dynamics of Kullback-Leibler Divergence Associated with Markov Chains:**
  In the spirit of above Lemma, we are naturally motivated to define Kullback-Leibler Divergence between the initial probability mass function ( of a DTMC/CTMC ) and the transient probability mass function at time 't' i.e.

  $$g(t) = D( \bar{\pi}(0) \, || \, \bar{\pi}(t) ).$$

  Since a homogenous, irreducible Markov chain reaches equilibrium probability distribution, it is clear that g(t) reaches an equilibrium value. We are currently studying the properties of g(t) [CoT].

- **Measures Defined on the Channel Matrix of a Discrete Memoryless Channel : Measures on arbitrary DTMCs**:
  Let the conditional probability mass functions specified by the rows of the channel matrix be given by

  $$\overline{q_1}, \overline{q_2}, \dots\dots, \overline{q_M} .$$

  Define the following two measures ( using the Kullback-Leibler divergence between any two rows of the channel matrix )

  $$M_1 = \underset{\substack{\{\overline{q_i}, \overline{q_j}; 1 \leq i,j \leq M\} \\ i \neq j}}{Min} D ( \overline{q_i} \, || \, \overline{q_j} )$$

$$M_2 = \underset{i \neq j}{Max\, D} \left( \bar{q}_i \, || \, \bar{q}_j \right) \{ \bar{q}_i , \bar{q}_j ; 1 \leq i,j \leq M \} .$$

- It is expected that these two measures ( which specify how differently the inputs of a Discrete Memoryless channel are garbled ) will be of utility in information theoretic investigations. In the case of DMC specified by a doubly stochastic matrix ( not necessarily a symmetric stochastic matrix ), similar measures are defined based on the columns of the channel matrix.

- Similar measures are defined on the state transition matrix ( a stochastic or doubly stochastic matrix ) of an arbitrary Discrete Time Markov Chain (DTMC) ( Using Minimum and Maximum possible Kullback-Leibler Divergence between the rows and/or columns )

## 5. Structured Graphs : Structured Markov Chains:

In the research area of applied probability, researchers realized that structured state transition matrices /generator matrices ( such as G/M/1-type, M/G/1-type structures ) lead to efficient computational forms for the equilibrium and /or transient probability distributions of DTMCs / CTMCs. It is evident that the structure in the state transition / generator matrices reflects in the corresponding structure of state transition diagram / state transition rate diagram. Thus, we arrive at structured directed graphs corresponding to structured Markov Chains. The efficient computational forms for equilibrium and transient probability distributions of structured Markov chains could have serious implications to spectral graph theory. We propose a new approach for imposing structure on the graphs corresponding to DTMCs / CTMCs. This approach is based on the vertex degree probability distribution defined in Section 3. We associate Shannon entropy with the vertex degree probability mass function and define it as the graph entropy. The following Lemma can easily be formally proved.

**Lemma 2:** The Graph entropy ( Shannon ) assumes the maximum possible value if and only if the degree of each vertex assumes the same value.

**Proof:** Follows from the basic properties of Shannon entropy ( maximization ) Q.E.D.

**Remark:** Ring connected graph, fully connected graph are the examples of graphs which assume maximum Shannon entropy ( based on the vertex degree probability mass function ).

Let us call structured graphs whose associated Shannon entropy is maximum as the Maximum-Entropic graphs. Thus ring connected graph, fully connected graph are the examples of such graphs ( always on any finite number of vertices ). Thus, the vertex degree matrix "D" of a maximum-entropic graph is given by

**D** = c **I,** for some constant integer value 'c'.

Hence the state transition matrix associated with the Discrete Time Markov Chain (DTMC) is given by

$$P = D^{-1} A = \frac{1}{c} A,$$

where **A** is the adjacency matrix. Thus, the transient probability distribution of such a DTMC is given by

$$\bar{\pi} \, (n+1) = \bar{\pi} \, (0) \, P^n = \frac{1}{c^n} \bar{\pi} \, (0) \, A^n \;\; for \;\; n \geq 0.$$

Also the equilibrium probability distribution is the maximum entropic distribution i.e.

$$\left[\frac{1}{M}, \quad \frac{1}{M}, \ldots \frac{1}{M}\right].$$

**Remark:** It should be noted that based on the connectivity structure, various possible graphs can be constructed ( on say a total of M vertices ). Among these graphs, maximum entropic graphs are distinguished. Also, one can easily arrive at minimum-entropic graphs on say M vertices. The adjacency matrix of a min-entropic graph is given by

$$A = \begin{bmatrix} 0 & 0 & \ldots\ldots & 0 & 1 \\ 0 & 0 & \ldots\ldots & 0 & 1 \\ \vdots & \vdots & \vdots & \vdots & \vdots \\ 0 & 0 & \ldots\ldots & 0 & 1 \\ 1 & 1 & \ldots\ldots & 1 & 0 \end{bmatrix}.$$

**Remark:** In the spirit of the discussion in sections 2 and 3, maximum entropic graphs and associated DTMCs/CTMCs can easily be studied ( for equilibrium and transient behavior ). Details are avoided for brevity.

**6. Conclusions**:

In this paper various ways of associating Discrete / Continuous Time Markov chains with weighted/unweighted , directed / undirected graphs are investigated. Also entropy dynamics associated with a structured Markov chain is investigated. Further certain structured graphs are considered and Markov chains are naturally associated with them.